\documentclass[reprint,superscriptaddress,amsmath,amssymb,longbibliography,twocolumn]{revtex4-2}

\usepackage{amsmath}
\usepackage{mathtools}
\usepackage{amssymb}
\usepackage{graphicx}
\usepackage{bm}
\usepackage[colorlinks, linkcolor=blue, citecolor=blue, urlcolor=blue,breaklinks=true]{hyperref}
\usepackage{color,dsfont,multirow,booktabs}
\usepackage{braket}
\usepackage{tabularx}
\usepackage{lipsum,booktabs}
\usepackage{ulem}
\usepackage{gensymb}

\begin{document}

\date{\today}

\title{Phase Dependent Quantum Optical Coherence Tomography  }

\author{Mayte Y. Li-Gomez}
\email{mayte.ligomez@ucalgary.ca}
\author{Taras Hrushevskyi}
\author{Kayla McArthur}
\affiliation{Institute for Quantum Science and Technology, and Department of Physics and Astronomy University of Calgary,
2500 University Drive NW, Calgary, Alberta T2N 1N4, Canada}

\author{Pablo Yepiz-Graciano}
\author{Alfred B. U'Ren}
\affiliation{Instituto de Ciencias
Nucleares, Universidad Nacional Autonoma de México,
04510 Ciudad de México, México}

\author{Shabir Barzanjeh}
\email{shabir.barzanjeh@ucalgary.ca}
\affiliation{Institute for Quantum Science and Technology, and Department of Physics and Astronomy University of Calgary,
2500 University Drive NW, Calgary, Alberta T2N 1N4, Canada}

\date{\today}

\date{\today}

\begin{abstract}
Entanglement is a key resource in quantum technologies, enhancing precision and resolution in imaging and sensing by leveraging the cross-correlation of photon pairs. This correlation enables precise time synchronization of photons reaching the photodetectors, effectively suppressing environmental noise and improving measurement accuracy. Building on this concept, we theoretically introduce and experimentally explore phase-dependent Quantum Optical Coherence Tomography. This technique employs phase-shifted entangled photon pairs for non-invasive morphological analysis of multilayered samples. 
We demonstrate that applying a phase shift to entangled photon pairs in the Hong-Ou-Mandel interferometer effectively eliminates artifacts (false patterns) caused by cross-reflections between different sample layers. This significantly improves the accuracy and reliability of the interferometric signal. The impact of this work extends to both the fundamental and practical domains. We show that phase-shifting entangled photon pairs in a Hong-Ou-Mandel interferometer can lead to tangible advancements in quantum sensing and probing. Practically, our method addresses a key challenge in Quantum Optical Coherence Tomography by eliminating artifacts, offering promising applications in biomedical imaging and material science.

\end{abstract}

\maketitle


\section*{Introduction}

Entanglement and quantum correlations are fundamental aspects of quantum mechanics, forming the foundation of many advanced quantum technologies. Entangled photon pairs exhibit strong correlations in properties such as polarization, energy, or momentum, even over long distances. These correlations enable transformative applications in quantum information \cite{RevModPhys.81.865}, quantum sensing \cite{Aslam2023, Pirandola2018, Barzanjeh, PhysRevLett.114.080503, GALLEGOTORROME}, computing \cite{Ladd2010, Harrow2017},  and communication \cite{couteau2023applications, Hanson}. Temporal and spectral correlations, in particular, play a crucial role in achieving precise measurements and enabling interference-based phenomena.

A prominent example of such quantum correlations is the Hong-Ou-Mandel (HOM) effect \cite{Bouchard_2021}, which occurs when two indistinguishable photons are directed into the input ports of a beam splitter. Quantum interference causes both photons to exit through the same output port rather than randomly splitting, a direct result of their indistinguishability. This effect, characterized by the "HOM dip" in coincidence detection, serves as a clear measure of photon indistinguishability and quantum correlations. The HOM effect is essential in quantum experiments and widely used in applications like quantum metrology and Quantum Optical Coherence Tomography (QOCT).

QOCT builds on the principles of entanglement to improve and extend the functionality of traditional Optical Coherence Tomography (OCT). Since its introduction in 1991 \cite{huang_OCT}, OCT has become an essential imaging tool in fields such as medicine and biology \cite{Everett_OCT_ophtalmology, Gabriele_3D-OCT, chopra_OCT_2020s}. Its ability to provide high-resolution, micrometer-scale images in a non-invasive manner has made it particularly indispensable in ophthalmology, where it is widely used for diagnosing and monitoring retinal conditions.
Using frequency-correlated photons, generated by spontaneous parametric down-conversion (SPDC), QOCT achieves a twofold improvement in axial resolution compared to OCT while maintaining immunity to group velocity dispersion \cite{Abouraddy_2002,Nasr_2003_expQOCT, Nasr_2004}. These properties make QOCT a promising next-generation imaging technology, capable of overcoming key limitations of its classical counterpart and opening new possibilities in medical diagnostics and biological research. \cite{Abouraddy_2002}. Recent works have highlighted the effectiveness of QOCT in different applications, including studies in dispersive media \cite{Nasr_2004} and biological tissue \cite{Nasr_2009_bio, Yepiz_Borja_2022_QOCM}. Furthermore, alternative methodologies, such as those that use phase conjugate cross-correlations \cite{Erkmen&Shapiro_PC-OCT} or chirped quasi-phase matching \cite{Carrasco_2004_chirpedQPM}, have been developed to achieve the enhanced resolution capabilities of QOCT while still maintaining high sensitivity. 

Despite significant progress and wide-ranging applications, QOCT still faces two main limitations: long acquisition times and the presence of artifacts. These artifacts originate from the quantum properties of the system and are caused by interference between photons reflected from different reflective layers within a sample. In particular, artifacts, along with recently identified echoes \cite{our_paper_2023}, appear as peaks or dips in the QOCT signal that do not correspond to actual physical interfaces within the sample. Additionally, for samples with more than two layers, artifacts are generated for every pair of real interfaces, making it increasingly challenging to extract accurate and meaningful information from complex structures.

Several approaches have been proposed to mitigate artifacts in QOCT. One of the earliest methods \cite{Abouraddy_2002} suggests removing artifacts by averaging interferograms obtained from multiple pump frequencies or by using a broad-bandwidth pump \cite{yepiz_2019}. Another strategy involves Fourier-domain QOCT, which helps address the issue of long acquisition times \cite{yepiz_2020_spectralHOM, kolenderska_vanholsbeeck_kolenderski_2020, kolenderska_szkulmowski_2021, kolenderska2025_arxiv}. However, this technique is constrained by spectral resolution and its reliance on highly sensitive single-photon detectors, which can introduce noise and reduce overall efficiency.

In this work, we introduce phase-dependent QOCT, a novel technique for artifact-free multi-layer sample probing by incorporating phase shifts into entangled photon pairs using electro-optic modulators. This method exploits the phase independence of the HOM effect while leveraging the fact that artifacts originate from self-photon interference, which is inherently phase-dependent. By modulating the phase of entangled photons in tomography, we can effectively control and even eliminate artifacts without affecting the HOM dips corresponding to different layers of the sample. This approach offers several advantages: It requires only minor modifications to conventional entangled-photon-based setups, as it can be implemented using standard electro-optic components, and it provides a straightforward yet highly effective solution for artifact management. By addressing a key limitation of QOCT, this technique enhances its feasibility for real-world applications, advancing the technology toward practical deployment and commercial viability.
 
\begin{figure}[t]
    \centering
    \includegraphics[width=0.9\linewidth]{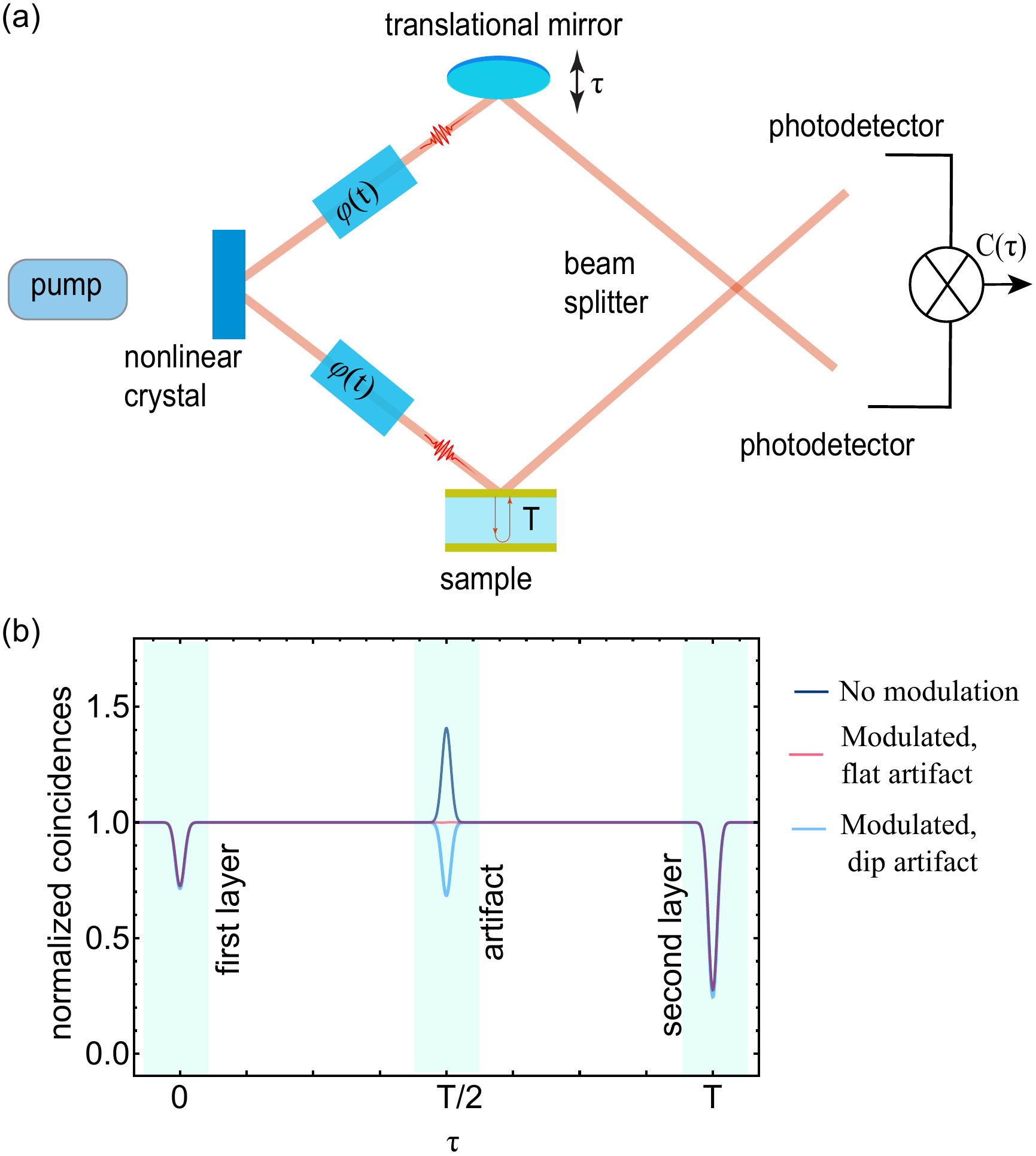}
    \caption{(a) 
    Schematic of Phase-Dependent QOCT. Entangled idler-signal photon pairs are generated via the SPDC process when a pump beam interacts with a nonlinear crystal. The generated photon pairs pass through two phase shifters $\phi_i(t)$ before being used to scan a sample. The idler photons then probe different layers of an unknown sample while the signal photons travel through a reference arm of adjustable length to match the path delay $\tau$ with the idler photons. The two photons are then recombined at a beamsplitter, where Hong-Ou-Mandel interference occurs. By measuring the coincidence rate $C(\tau)$, an interferogram is obtained, revealing information about the internal interfaces of the sample. (b) Full QOCT coincidence interferogram displaying a HOM dip for each reflective layer in a two-layer sample, along with a peak artifact. The artifact's amplitude is controlled by introducing phase modulation through variations in $\beta_i$ ($i=1,2$), while the phase change does not impact the HOM dips.}
\label{fig: basicQOCT_phaseMod_interferogram}
\end{figure}

\section{Theory of phase-dependent QOCT}

In this section, we introduce a theoretical framework for phase-dependent QOCT and demonstrate that applying phase shifts to entangled photons using a phase-modulated source results in a completely artifact-free QOCT. To explore this, the discussion is divided into two parts: First, we develop a theoretical model in which entangled pairs generated by an entangled source undergo a phase shift, and then we build upon this result to introduce the concept of phase-dependent QOCT.
\subsection{Phase modulation of entangled photon-pairs}

We start by introducing the initial quantum state of the entangled photon pairs generated via spontaneous parametric down-conversion (SPDC):

\begin{equation}
\ket{\psi}=\ket{0}+\eta \int \int d\omega_1 d\omega_2 f(\omega_1,\omega_2) \ket{\omega_1}\ket{\omega_2},
\end{equation}
where $f(\omega_1,\omega_2)$ is the joint spectral amplitude (JSA) of the single-photons as determined by the phase-matching conditions of the down conversion process having efficiency $\eta$. Note that $S(\omega_1,\omega_2)\equiv\lvert f(\omega_1,\omega_2) \rvert^{2} $ represents the joint spectral intensity, which describes the probability of detecting a signal photon with frequency $\omega_1$ and an idler photon with frequency $\omega_2$. This can be approximated as a Gaussian function \cite{our_paper_2023}
\begin{eqnarray}
\label{eq_joint_spectral_intensity}
S(\omega_1,\omega_2)=\frac{4}{\pi \sigma_a \sigma_d} 
&\exp&\left[-2\left(\frac{\omega_1+\omega_2-2\omega_0}{\sigma_d} \right)^2 \right]\nonumber\\
\times &\exp&\left[-2\left(\frac{\omega_1-\omega_2}{\sigma_a} \right)^2 \right],
\end{eqnarray}
where $\omega_0$ is the pump frequency and $\sigma_a$ ($\sigma_d$) represents the bandwidth along the antidiagonal (diagonal) in the ${\omega_1, \omega_2}$ frequency space. In general, $\sigma_a$ can be experimentally adjusted using a bandpass filter applied to the photon pairs, whereas $\sigma_d$ can be controlled through the pump bandwidth.

We now analyze the effect of applying an external phase to the biphoton state. Assume that an external phase $\phi$ is applied to each photon in the entangled pair. This is implemented in practice by placing two EOMs along the propagation paths of the biphotons. A sinusoidal electric field drives each EOM with a frequency $\Omega$, where the amplitude of the field proportionally modulates the phase of the transmitted optical field. The additional phase is given by $ \phi(t)=  \beta \sin({\Omega t+\theta})$, 
where $\beta = \frac{V}{V_{\pi}} \pi$ is the modulation index describing the phase deviation of the modulated signal away from the unmodulated one, $V$ is the voltage amplitude of the signal, and $V_{\pi}$ is the half-wave voltage of the EOM. As a result, the output field of the EOM can be expressed as 
\begin{equation}
      e^{i \phi(t)} = \sum_{m=-\infty} ^{\infty} J_{m} (\beta) e^{i m (\Omega t +\theta)},
\label{eq: modulation_Jacobi-Anger}
\end{equation}
where $J_{m} (x)$ is the Bessel functions of the first kind of order $m$. Given that the phases are time-dependent, it is practical to analyze the system in the temporal domain, where the JSA of the SPDC photons can be represented as
\begin{equation}
    f (\omega_{1}, \omega_{2}) = \int \int dt_{1} dt_{2} f(t_{1},t_{2}) e^{- i \omega_{1} t_{1}} e^{- i \omega_{2} t_{2}},
\label{eq: JSI_ft}
\end{equation}
where $f(t_{1},t_{2})$ is the joint temporal amplitude of the biphotons. As a result, the total JSA of the photon pairs after the phase modulation becomes

\begin{eqnarray}
    &&f_{PM}(t_{1}, t_{2}) = f (t_{1}, t_{2}) \, \times  \\ 
    &&\sum_{m=-\infty} ^{\infty} J_{m} (\beta_{1}) e^{i m (\Omega_{1} t_{1} +\theta_{1})} \sum_{n=-\infty} ^{\infty} J_{n} (\beta_{2}) e^{i n (\Omega_{2} t_{2} +\theta_{2})}, \nonumber
    \label{eq:JSI_JA}
\end{eqnarray}

In the frequency domain, the phase-shifted JSA can be written as

    \begin{eqnarray} \label{eq:jsa_pm0}
           f_{PM}(\omega_{1}, \omega_{2})&=& \int \int dt_{1} dt_{2} f_{PM}(t_{1}, t_{2}) e^{-i \omega_{1} t_{1}} e^{-i \omega_{2} t_{2}}\\\nonumber&=&
 \sum_{m=-\infty}^{\infty}\sum_{n=-\infty}^{\infty} f_{mn}(\omega_1,\omega_2),
\end{eqnarray}

where 
\begin{eqnarray}
  f_{mn}(\omega_1,\omega_2)&=& f(\omega_1-m\Omega_1,\omega_2-n\Omega_2) \times \nonumber \\
&& \qquad \qquad J_m(\beta_1)J_n(\beta_2) e^{i(m\theta_1+n\theta_2)}.
\end{eqnarray}

Equation (\ref{eq:jsa_pm0}) describes the appearance of multiple sidebands in the spectral space that compose a network of sub-JSAs, $f_{mn}$, which are separated by integer multiples of the external driving frequency $m \Omega_1$ and $n \Omega_2$.

\subsection{Phase-dependent QOCT}

QOCT takes advantage of HOM interference to reconstruct the internal axial structure of multilayered samples. Pairs of entangled photons, often generated by type-I or type-II SPDC, serve as inputs for the HOM interferomenter. The signal photon is sent to the reference arm with a controllable delay $\tau$, while the idler photon is sent to the sample arm in which light is reflected back from each of the reflective surfaces within the sample. The two photons are then recombined and interfered in a beamsplitter. The output of their interference $C(\tau)$ is recorded as a function of the temporal delay between the reference and sample arm. A basic schematic of a QOCT with phase modulation on the entangled pairs is shown in Figure \ref{fig: basicQOCT_phaseMod_interferogram}a.  Due to the HOM effect, a dip appears in the interferogram when the optical path lengths of the signal and idler photons are equal. 
The distance between the HOM dips in the interferogram allows information concerning the sample's internal morphology to be extracted.

The output of the HOM interferometer is a second-order correlation measurement, which can be mathematically described by the coincidence interferogram function \cite{yepiz_2019}

\begin{align}
     C(\tau)=\frac{N_0}{4}\int \int \mathrm{d\omega_1 d\omega_2}
        |f_{PM}(\omega_1,\omega_2) \, H(\omega_2) \nonumber  \\-f_{PM}(\omega_2,\omega_1) \, H(\omega_1) \, e^{i(\omega_2-\omega_1)\tau}|^2,
    \label{eq: C(tau)}
\end{align}
where \( N_{0} \) represents the background coincidence count rate, and \( H(\omega) \) represents the Sample Transfer Function. For a two-layer sample, this function is given by $ H(\omega) = r_1 + r_2 e^{i\omega T}$, where \( r_1^2 \) and \( r_2^2 \) correspond to the reflectivities of the two layers, and \( T \) represents the propagation time between them. By expanding each term, we obtain
\begin{equation}
    C(\tau) = \frac{N_{0}}{4} \Big[G_0 -G(\tau)\Big],
\end{equation}
where
\begin{align}
     G_0 &= \int \int \mathrm{d\omega_{1}} \mathrm{d\omega_{2}} \Big[ \Big|f_{PM}(\omega_{1}, \omega_{2})H(\omega_{2})\Big|^{2} \nonumber\\
     &+ \Big|f_{PM}(\omega_{2}, \omega_{1})H(\omega_{1})\Big|^{2}  \Big],
\end{align}
is the self-interference terms which are independent of $\tau$, and the cross-interference terms \cite{Abouraddy_2002, yepiz_2019}
\begin{align}\label{Gtatu}
    G(\tau) = \int \int \mathrm{d\omega_{1}} \mathrm{d\omega_{2}} \Big[&f_{PM}(\omega_{1}, \omega_{2}) f^{*}_{PM}(\omega_{2}, \omega_{1})\\ & H^{*}(\omega_{1}) H(\omega_{2}) e^{-i (\omega_{2}-\omega_{1})\tau}+C.C\Big],\nonumber
\end{align}
which contains information about the light reflected back from each of the internal layers of the sample and, therefore, about its axial morphology. By substituting Eq. (\ref{eq:jsa_pm0}) in Eq. (\ref{Gtatu}) and after a few algebraic simplifications we obtain 

\begin{widetext}
   \begin{multline}
    G(\tau)= 2 \sum_{mn}\sum_{m'n'} \Lambda_{mn}\Lambda_{m'n'} \, e^{-\frac{(\Delta^+_{mn}-\Delta^+_{m'n'})^2}{2\sigma_d^2}}e^{-\frac{(\Delta^-_{mn}+\Delta^-_{m'n'})^2}{2\sigma_a^2}}
    \Bigg[ r_1^2 \kappa(0)+r_2^2 \kappa\left(T\right) + \\ 
        r_1 r_2 \, e^{-\frac{\sigma_{d}^{2}}{8} (\frac{T}{2})^{2}} \cos\left( \omega_{0} T + \frac{(\Delta_{m,n}^{+} + \Delta_{m^{\prime},n^{\prime}}^{+})}{2} \frac{T}{2}\right) \kappa\left(\frac{T}{2}\right) \Bigg],
\label{eq: artifact_theory}
\end{multline}  
\end{widetext}
where we have defined $\Lambda_{mn}\equiv J_{m}(\beta_{1}) J_{n}(\beta_{2})$, $
   \Delta^{\pm}_{m,n} \equiv m \Omega_{1} \pm n \Omega_{2},$ and
\begin{eqnarray}
      \kappa&&(\mathcal{T})=\nonumber e^{-\frac{\sigma_{a}^{2}}{8} (\mathcal{T}-\tau)^{2}} \times \\
      && \cos{\left( \frac{(\Delta_{m,n}^{-} - \Delta_{m^{\prime},n^{\prime}}^{-})}{2}  (\mathcal{T}-\tau) - \Theta^{m,n}_{m^{\prime},n^{\prime}} \right)},
    \label{eq: mod_function}
\end{eqnarray}
with $\Theta^{m,n}_{m^{\prime},n^{\prime}} \equiv (m-m^{\prime}) \theta_{1} + (n-n^{\prime}) \theta_{2}$. The first two terms inside the brackets in Eq. ({\ref{eq: artifact_theory}}) correspond to the first and second HOM dips, associated with the first layer at $\tau=0$ and the second layer at $\tau=T$, respectively. Meanwhile, the third term represents the artifact that appears at the midpoint of the sample, occurring at
$\tau=T/2$. Eq. {(\ref{eq: artifact_theory})} is generally an interesting yet somewhat complex expression. The response of the system has been plotted in Fig. \ref{fig: basicQOCT_phaseMod_interferogram}b which shows the normalized interferogram \( \Gamma(\tau) = 1 - G(\tau)/G_0 \) for different values of $\beta = \frac{V}{V_{\pi}} \pi$ by changing the applied RF voltages. This figure illustrates that while the HOM dips remain unchanged, the artifacts can be fully inverted from a peak to a dip or completely suppressed by appropriately tuning the phases. This behavior can be partially understood by focusing only on the diagonal terms in the summation, where \( n = n' \) and \( m = m' \). Under these conditions, $
\kappa(\mathcal{T}) = e^{-\frac{\sigma_{a}^{2}}{8} (\mathcal{T}-\tau)^{2}}
$ and $
\Theta^{m,n}_{m',n'} = 0,$ thus simplifying Eq. {(\ref{eq: artifact_theory})} to:

\begin{eqnarray}
    G(\tau)&=& 2 \sum_{mn} \Lambda_{mn}^2 \, e^{-\frac{2(\Delta^-_{mn})^2}{\sigma_a^2}}
    \Bigg[ r_1^2 \kappa(0)+r_2^2 \kappa\left(T\right) \\
       &+& r_1 r_2 \, e^{-\frac{(\sigma_{d}T)^{2}}{32}} \mathrm{cos}\left( \omega_{0} T +\frac{T\Delta_{m,n}^{+}}{2}\right) \kappa\left(\frac{T}{2}\right) \Bigg],\nonumber
\end{eqnarray}  
The above equation explicitly demonstrates that the first and second terms inside the brackets remain unaffected by the phase injected into the photon pairs, whereas the third term depends on $
   \Delta^{+}_{m,n} \equiv m \Omega_{1}+ n \Omega_{2}$. In the following section, we will experimentally investigate the phase-dependent QOCT.

\begin{figure*}[t]
    \centering
    \includegraphics[width=0.85\linewidth]{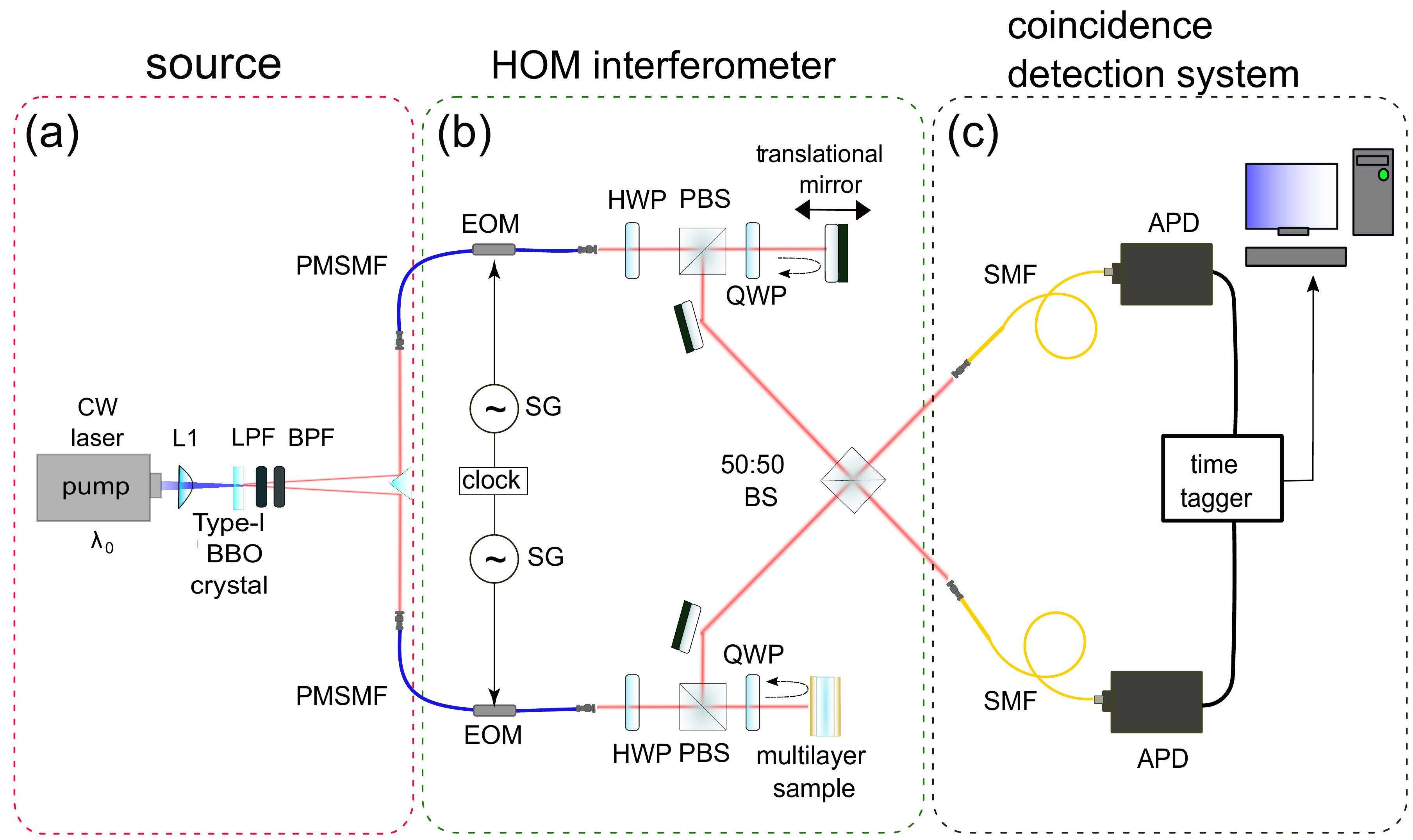}
    \caption{\textbf{Phase-dependent Quantum Optical Coherence Tomography Setup.} The system consists of a Spontaneous Parametric Down-Conversion (SPDC) entangled photon pair source, a Hong-Ou-Mandel (HOM) interferometer, and a coincidence detection system. A tunable continuous-wave (CW) laser with a wavelength of $\lambda_{0} = 404.5$ nm pumps a type-I $\beta$-Barium Borate (BBO) nonlinear crystal, generating entangled idler-signal photon pairs. The generated entangled pairs pass through two identical electro-optic modulators (EOMs), both driven by locked microwave signal generators (SG). The phase-shifted photon pairs are directed into the HOM interferometer, where they are recombined and subsequently detected using avalanche photodiodes (APDs). Key optical components include polarization-maintaining single-mode fibers (PMSMFs), a polarization beam splitter (PBS), a quarter-wave plate (QWP), a beam splitter (BS), low/band-pass filters (L/BPF), a single-mode fiber (SMF), and a half-wave plate (HWP).}
    \label{fig:phaseMod_setup}
\end{figure*}

\section{Experimental Setup}
Figure \ref{fig:phaseMod_setup} shows the experimental setup for phase-dependent QOCT. In essence, the setup consists of the source of entangled pairs, the modified HOM interferometer, and the coincidence detection system. 
The system is pumped with a tunable continuous-wave (CW) laser, emitting light with a pump wavelength centered at $\lambda_{0} = 404.5$ nm and set to an average power of 60 mW. The pump polarization and beam waist diameter are controlled with a half-waveplate (HWP) and a planoconvex lens (L1, $f= 1000$ mm), respectively, that have been set to maximize transmission and focus the beam to have a diameter of $300\, \mu$m. The pump feeds a 2-mm long $\beta$-Barium Borate (BBO) crystal, which generates pairs of entangled photons through type-I SPDC. The signal and idler photons are emitted noncollinearly at an exit angle of $3^{\circ}$ with respect to the pump axis. The output is filtered with a long-pass filter (LPF) with cut-off wavelength $\lambda=500$ nm which blocks the pump and a spectral  bandpass filter (BPF) centered at $800$ nm, with a 40 nm width. The signal and idler photons are separated using a right-angle prism mirror that leads each of the twins into polarization-maintaining single-mode fibers (PMSMFs), which provide spatial mode filtering and ensure that the arms of the interferometer are fed with Gaussian modes. The output of the PMSMFs feeds the sample and reference arms of the interferometer. 

The phase of each twin photon is independently modulated using EOMs placed at the output of the PMSMFs. These EOMs are low-loss lithium niobate phase modulators designed to operate around 795 nm. Each modulator is driven by a signal function generator (SFG), which provides a signal with adjustable amplitude, driving RF frequency, and phase difference relative to a common rubidium clock that serves as a reference. The SFGs have a maximum output frequency of 12.7 GHz and an output voltage of 7.24~Vpp (22 dBm), corresponding to a range of $\beta_{1} \leq 5.42$ and $\beta_{2} \leq 4.48$ since $V_{\pi_{1}} = 3.08\, \mathrm{V}$ and $V_{\pi_{2}} = 3.73\, \mathrm{V}$.  The values of $\beta$ include a correction in input power due to cable losses. 

The signal photon is directed to the reference arm, where a temporal delay is introduced using a system consisting of a polarization beam splitter (PBS), a quarter-wave plate (QWP) that ensures the correct polarization for interference, and a translational mirror mounted on a motorized precision linear stage with a micrometer step size. The temporal delay $\tau$ is controlled by adjusting the path-length difference between the signal and idler photons.

In the sample arm, the idler photon probes a multilayered sample composed of two reflective interfaces. The sample used in our experiment is 2 mm-thick glass coated at the front and back with a thin gold layer via electron beam deposition. The reflectivities of the front and back layers (along the direction of light propagation) are $36\%$ and $95\%$, respectively. The signal and idler photons are then recombined at a 50:50 beam splitter (BS), which forms the core of the HOM interferometer. The output beams at the beam splitter’s exit ports are coupled into avalanche photodetectors (APDs). The normalized coincidence interferogram $\Gamma(\tau)$, resulting from HOM interference, is measured using a time controller unit with a fixed coincidence window $2$ ns.

\section*{Results}

The impact of phase modulation on artifacts in the QOCT trace is assessed by varying the phase of the entangled photons through adjustments to the EOMs' voltage or applied power, phase, and frequency. The resulting coincidence interferograms are recorded by independently varying EOMs parameters and performing a scan over the temporal delay $\tau$ (in ps), corresponding to the optical path length (in $\mathrm
{\mu m}$) traveled by light through the sample.  A HOM dip appears in the event that the path length difference between the reference and sample arm is zero ($\tau = 0$); that is, when the idler photon is reflected back from a reflective interface within the sample. While the entire sample length is scannable, our focus is on the region where the artifact appears. Additionally, the front and back surfaces are examined to confirm that the visibility of their respective HOM dips remains unchanged by phase modulation (see Supplementary Information). This is because the size of the sidebands, determined by the applied frequency and power, is much smaller than the bandwidth of the biphoton state.

\begin{figure}[t]
    \centering
    \includegraphics[width=0.95\linewidth]{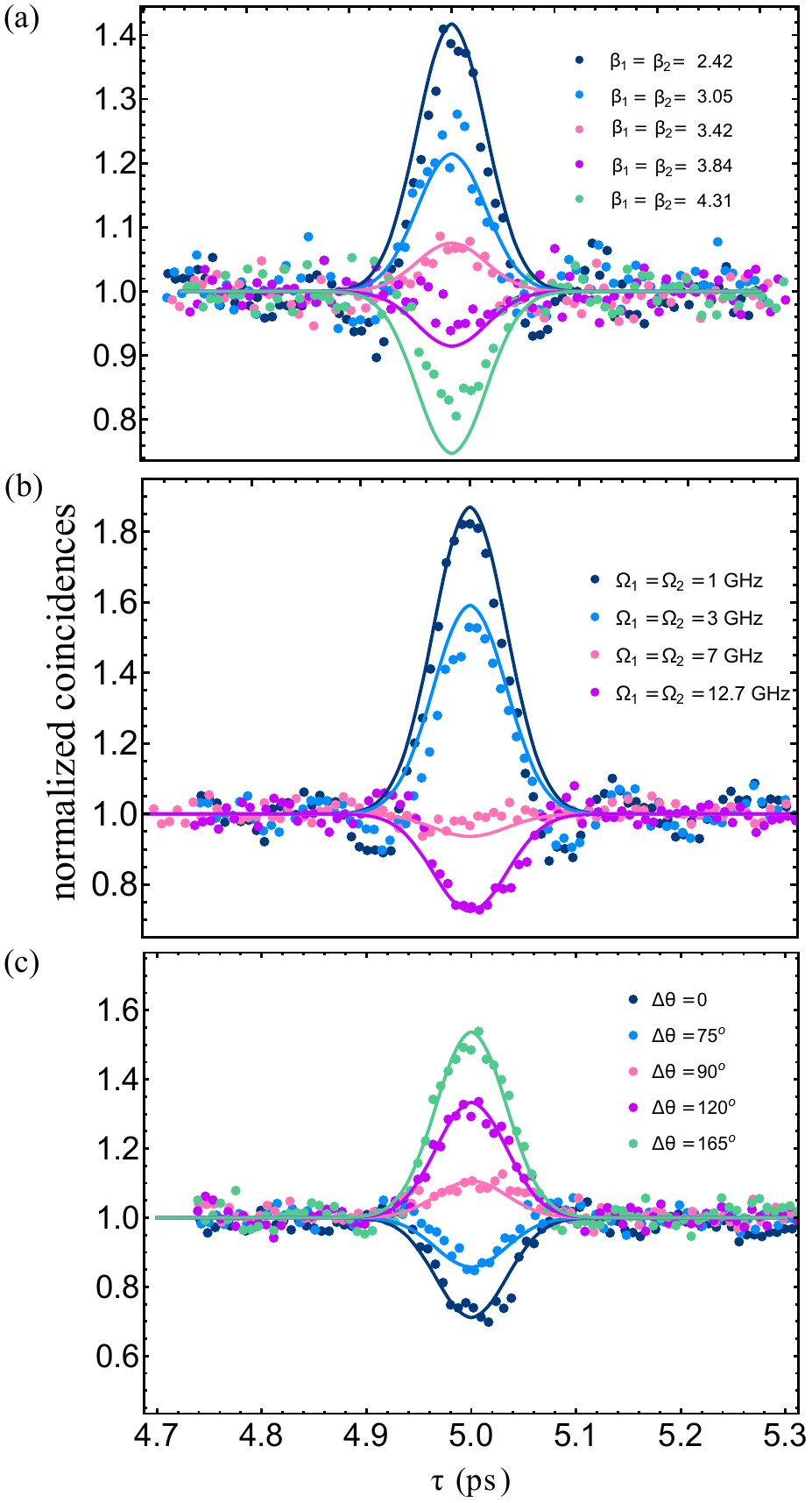}
    \caption{Effect of Modulation Parameters on Artifact Visibility. The visibility of the artifact, measured in normalized coincidences, varies as a function of: (a) the voltage amplitudes $\beta_1$ and $\beta_2$, while keeping the driving frequencies fixed at $\Omega_{1}/ 2\pi=\Omega_{2}/2\pi= 12.7$ GHz and the phase difference at $\Delta\theta = 0$; (b) the driving frequencies $\Omega_1=\Omega_2$, with a fixed phase difference of $\Delta\theta = 0$ and maximum voltage applied to each EOM; and (c) the phase difference between the microwave signals driving the EOMs, when each EOM is driven at maximum voltage at a frequency of 12.7~GHz. Here, the experimental data (dots) have been fitted using our theoretical model (solid line).  }
    \label{fig: fitted_plots_PFD}
\end{figure}

Figure \ref{fig: fitted_plots_PFD} illustrates the effect of phase modulation on the artifact for various parameters of both EOMs in the idler (EOM 1) and signal (EOM 2) paths, including the applied voltages ($\beta_i$), RF driving frequency $\Omega_i$, and phase difference between the two sources $\Delta\theta=\theta_2-\theta_1$, with $i=1,2$. Although this experiment uses two EOMs, it should be noted that our protocol does not require both to manipulate the phase. A single EOM can also achieve the desired effect. 
In each measurement set, we isolate one specific variable and systematically scan its entire range while keeping all other variables constant. The amplitude of the artifact is reported in normalized coincidences \( \Gamma(\tau) = 1 - G(\tau)/G_0 \).

The voltage scans in Fig. \ref{fig: fitted_plots_PFD}a are obtained by fixing the driving frequencies at $\Omega_{1}/ 2\pi=\Omega_{2}/2\pi= 12.7$ GHz and the phase difference $\Delta\theta = 0$. The values of frequency $\Omega_{i}$ and $\Delta\theta$ were specifically selected to achieve the largest change in the amplitude of the artifact and demonstrate the effect of modulation in detail. The polarity of the artifact becomes more negative for higher values of $\beta$. Noticeable changes appearing at $\beta_{1}=2.16, \beta_{2}= 1.79$, corresponding to peak-to-peak voltage of 2.12~V, the voltage was then gradually increased to the maximum power output of the signal generators at $\beta_{1}=5.42, \beta_{2}=4.48$. This led to a full flip on the artifact from peak to dip. 

The effect of varying the driving frequencies of both EOMs is shown on Fig. \ref{fig: fitted_plots_PFD}b. The microwave source frequencies for both EOMs are scanned symmetrically, \(\Omega_{1} = \Omega_{2}\), over the range of 1 to 12.7~GHz, while maintaining a fixed phase difference of \(\Delta\theta = 0\) and applying maximum voltage to each EOM. As seen, the artifact transitions from a peak to a dip and is effectively eliminated at \(\Omega_{1}/2\pi = 7\)~GHz. It is worth mentioning that this frequency is particular to the thickness of the sample used in our experiment but will vary for different values of T.

Fig. \ref{fig: fitted_plots_PFD}c illustrates the impact of the phase shift between the EOMs when driven at maximum voltage and a frequency of 12.7~GHz. As with the previous two configurations, the amplitude of the artifact can be controlled, in this case, by adjusting the phase difference $\Delta \theta$. Interestingly, the phase modulation effect vanishes as the phase difference $\Delta \theta$ approaches $\pi$ (see also SI Fig. \ref{fig:Supplementary phase scans plot}b for 9~GHz), suggesting that the influence of the EOMs cancel each other. The solid lines in Fig. \ref{fig: fitted_plots_PFD} (a)-(c) show the theoretical fits using Eq. (\ref{eq: artifact_theory}).

\begin{figure}[t!]
    \centering
    \includegraphics[width=\linewidth]{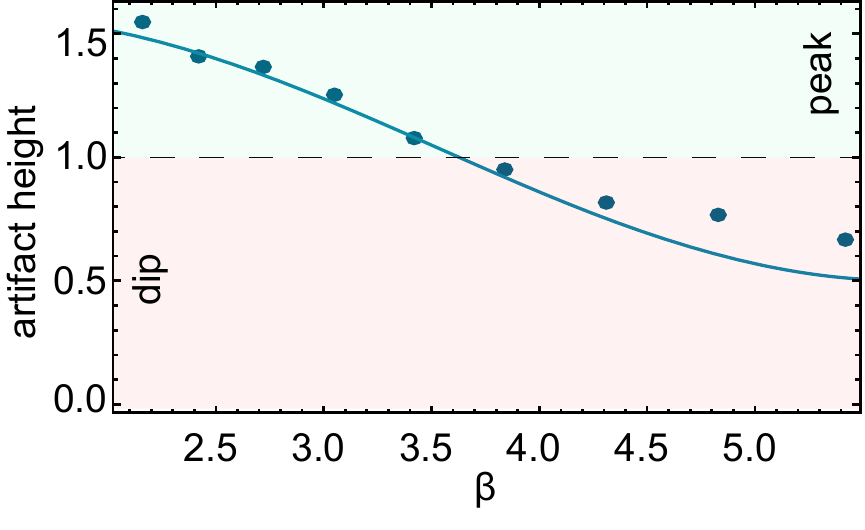}
    \caption{The artifacts transition from a peak to zero and then to a dip as the modulation depth is adjusted, with $\beta = \beta_1 = \beta_2$ for the driving frequencies fixed at $\Omega_{1}/ 2\pi=\Omega_{2}/2\pi= 12.7$ GHz and the phase difference at $\Delta\theta = 0$. Here, the experimental data (dots) have been fitted using our theoretical model (solid line).}
    \label{fig4}
\end{figure}

 Finally, the results of the fitting using model based on Eq. (\ref{eq: artifact_theory}) for different values of $\beta$ are shown in Fig. \ref{fig4}. The impact on the visibility of the artifact ($\tau= T/2$) is measured for increasing values of $\beta$.  The height of the artifact transitions completely from a peak to zero and then to a dip by adjusting $\beta=\beta_1=\beta_2$. In this case, the voltage scans are recorded while keeping the driving frequencies fixed at $\Omega_{1}/ 2\pi=\Omega_{2}/2\pi= 12.7$ GHz and the phase difference $\Delta\theta = 0$. The experimental data (dots) have been fitted using our theoretical model (solid line).


\section*{Discussion and Conclusion}
In this work, we have demonstrated that introducing phase shifts to entangled photon pairs unlocks new possibilities in quantum imaging and precision probing, particularly enabling phase-dependent QOCT. Our experiment presents a straightforward yet effective approach to suppressing artifacts in QOCT signals by leveraging phase modulation of a quantum-entangled light source. Using EOMs, we introduced controlled phase shifts to the biphoton state, actively shaping the artifact's interference in the HOM pattern. This method provides a practical and easily implementable solution for enhancing QOCT performance with minimal modifications to existing experimental setups.

Our results indicate that the visibility of artifacts is primarily dictated by the phase difference between the independent driving signals applied to the EOMs. Additionally, variations in other parameters, such as the voltage and frequency of the EOMs, yield similar effects on artifact suppression. Although phase-dependent QOCT effectively eliminates artifacts, it still faces the challenge of long acquisition times. Nevertheless, our technique introduces a promising approach to advancing quantum probing and expanding the capabilities of quantum-enhanced measurement technologies. Beyond QOCT, our method establishes a novel strategy for quantum state manipulation, with potential applications in quantum sensing, metrology, and information processing. The ability to precisely control the phase of entangled photons could pave the way for the development of new types of sensors that utilize phase-dependent entangled photons to detect and estimate small phase shifts with high precision.

\begin{acknowledgments}
We thank Daniel Oblak and Sourabh Kumar for helping to build the phase modulation part of the experiment. S.B. acknowledges funding by the Natural Sciences and Engineering Research Council of Canada (NSERC) through its Discovery Grant and Quantum Alliance Grant, funding and advisory support provided by Alberta Innovates (AI) through the Accelerating Innovations into CarE (AICE) -- Concepts Program, support from Alberta Innovates and NSERC through Advance Grant project, and Alliance Quantum Consortium. This project is funded [in part] by the Government of Canada. Ce projet est financé [en partie] par le gouvernement du Canada. The authors would like to thank Dr. Daniel Oblak and Dr. Lindsay LeBlanc for generously lending us their electro-optical modulators, which were essential to the realization of this experiment. Lastly, the authors are deeply grateful to Dr. Erhan Saglamyurek for his insightful contributions, which greatly enriched the initial stage of this research. 
\end{acknowledgments}


\clearpage

\onecolumngrid
\appendix
\begin{center}
\textbf{\large Supplementary Materials}
\end{center}

\subsection{Full QOCT interferogram of the 2mm sample}

\begin{figure}[t]
    \centering
    \includegraphics[width=1\linewidth]{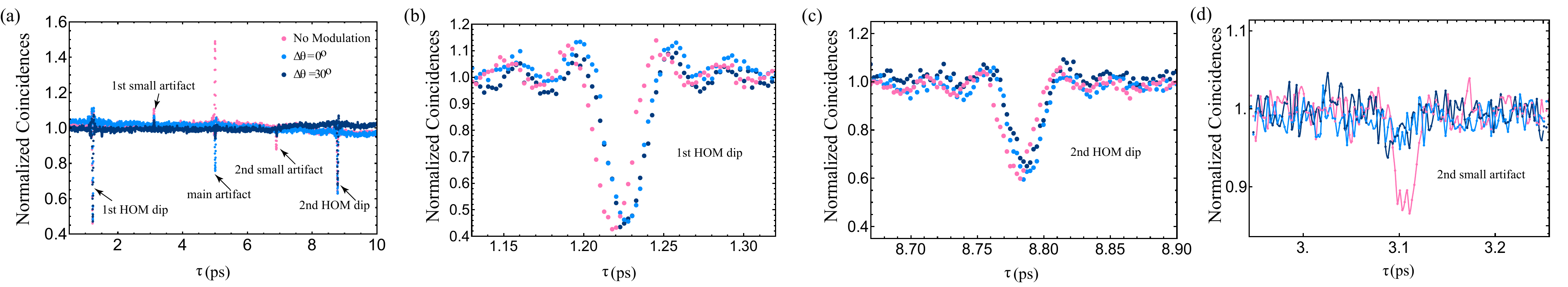}
    \caption{ (a) Full QOCT inteferograms of a 2mm sample, with and without modulation. The interferogram shows two HOM dips corresponding to real interfaces and one artifact at the midpoint between them. Smaller artifacts appear due to an air gap in the sample acting as a layer. (b) A close up of the first HOM dip, (c) a close up of the second HOM dip, (d) Close up of the second small artifact in the experimental scan.  }
    \label{fig:Supplementary Experimental plot}
\end{figure}

Figure \ref{fig:Supplementary Experimental plot} presents the complete interferogram of a sample of three interfaces, conducted to demonstrate the proof of principle for phase modulated QOCT and to confirm that phase modulation does not affect the visibility of the HOM dips. Figures \ref{fig:Supplementary Experimental plot} (b) and (c) present a close-up view of the first and the last HOM dips which show no change in their visibility, aligning with the theoretical model presented in the Main text.

The sample was fabricated by stacking two 1 mm-thick glass cover slips and depositing a thin gold film on their outer surfaces. The separation between the cover slips creates a thin air gap, introducing an additional layer with a reflective index of $r=0.04$. As a result, three artifacts appear in the interferogram: the main one at the midpoint between the front and back gold depositions and two smaller artifacts at $\tau=3.1$~ps and $\tau=7.1$~ps. While this behavior is consistent with our previous investigation of multilayered samples, the appearance of these smaller artifacts is not accounted for in the model presented in this work since it assumes a two-layer sample.

Nevertheless, these small artifacts, together with the main one, are removed by modulating the phase.
Figure \ref{fig:Supplementary Experimental plot} (d), shows a closer look at the first small artifact appearing at $\tau=3.1$ ps in the experimental scans. During the scans, the phase $\Delta \theta$ was varied, while the voltage and driving frequency were constant $\Omega_{1}/2\pi = \Omega_{2}/2\pi= 12.7$~GHz.

\subsection{Artifact removal with different drive frequencies}
 
\begin{figure}[t]
    \centering
    \includegraphics[width=1\linewidth]{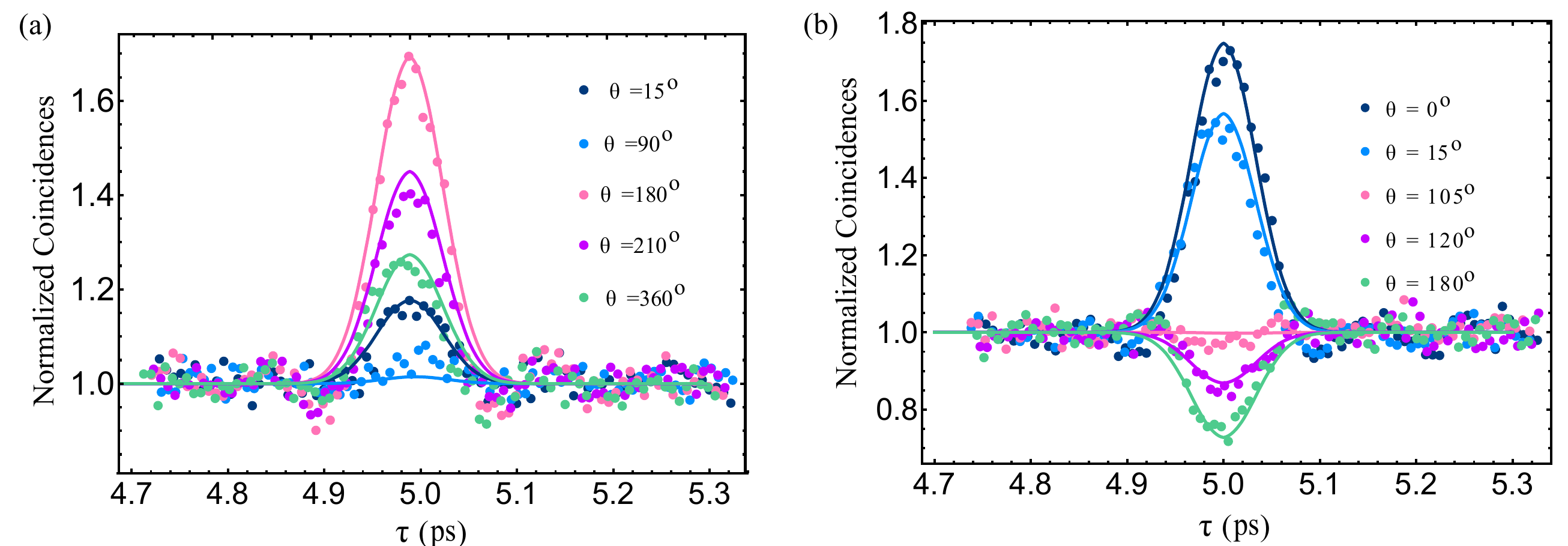}
    \caption{The visibility of the artifact, measured in normalized coincidences, changing as a result of varying the phase of the EOM with a) both EOMs fixed at $\Omega_{1}/2\pi= \Omega_{1}/2\pi= 6$~GHz, and b) both EOMs fixed at $\Omega_{1}/2\pi= \Omega_{1}/2\pi= 9$~GHz. }
    \label{fig:Supplementary phase scans plot}
\end{figure}

Figure \ref{fig:Supplementary phase scans plot} shows scans of $\Delta\theta$ at different phases $\Delta\theta = 15\degree, 90\degree, 180\degree, 210\degree, 360\degree$. The driving signal was set to a constant voltage of 7.24~V and constnat driving frequencies $\Omega_{1}/2\pi= \Omega_{1}/2\pi= 6$~GHz (Figure \ref{fig:Supplementary phase scans plot}a) and $\Omega_{1}/2\pi= \Omega_{1}/2\pi= 9$~GHz (Figure \ref{fig:Supplementary phase scans plot}b).

\end{document}